# A study on Modern Messaging Systems- Kafka, RabbitMQ and NATS Streaming

Sharvari T
*Department of Electronics and Communication*
*R.V. College of Engineering*
Bengaluru, India
sharvarit.ec15@rvce.edu.in

Sowmya Nag K
*Department of Electronics and Communication*
*R.V. College of Engineering*
Bengaluru, India
sowmyanagk@rvce.edu.in

*Abstract*—Distributed messaging systems form the core of big data streaming, cloud native applications and microservice architecture. With real-time critical applications there is a growing need for well-built messaging platform that is scalable, fault tolerant and has low latency. There are multiple modern messaging systems that have come up in the recent past, all with their own pros and cons. This has become problematic for the industry to decide which messaging system is the most suitable for a specific application. An in-depth study is required to decide which features of a messaging system meet the needs of the application. This survey paper outlines the modern messaging technologies and delves deep on three popular publisher/subscriber systems- Apache Kafka, RabbitMQ and NATS Streaming. The paper provides information about messaging systems, the use cases, similarities and differences of features to facilitate users to make an informed decision and also pave way for future research and development.

## I. INTRODUCTION

The rise of the internet has been coupled with the exponential consumption of data. Distributed systems are now widespread and have more than thousands of entities. They are constantly evolving. This underlines the need for cross platform communication method that can adapt to different protocols and is dynamic in nature. Point to point and synchronous communication do not adapt to the dynamic nature of applications. A scalable and loosely coupled system such as the publish-subscribe system [1] is well suited for the current market needs. Another similar system are message queues. Real-time data analytics, website tracking, logging and recent boom in IoT devices [2] has increased the demand for fault tolerant, highly available messaging systems. They form the middleware infrastructure for big data streaming, microservices and cloud-based applications.

### A. Publish/Subscribe Paradigm

Publish-subscribe pattern corresponds to a mechanism where in producers publish messages that are grouped into categories and consumers subscribe to categories which they are interested in. Fig.1 shows the overview of the pub-sub system with the publishers, subscribers and broker(server). The broker ensures the subscribers receive data they are interested in. As mentioned in [1] the strength of the system lies in

Time Decoupling: The publishers and subscribers need not be active at the same time.

Space Decoupling: The publishers and subscribers do not know each other neither do they know how many subscribers/publishers are there respectively.

Synchronization Decoupling: The publishers can publish message which can be retrieved by subscribers at any time. The production and consumption does not happen in a synchronous manner.

There are various modifications of this loosely coupled system present in the industry according to use cases and research developments. They are used for fraud detection, surveillance, air traffic control and algorithmic trading. In [3] pub/sub system is implemented in Microsoft Azure for cloud applications.

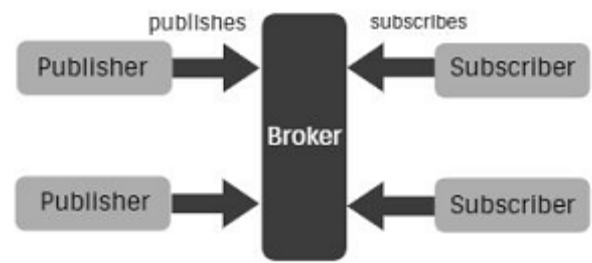

Fig.1 Publisher-Subscriber System

### B. Messaging Queue Paradigm

Producers send messages to queues, from where the consumers consume the data in order. Each message can be consumed only once. This is asynchronous in a sense that producers and consumers need not simultaneously interact with the message queue. The messages are stored in queue until it is retrieved. This paradigm is also known as point to point communication. Well-known messaging queue platforms are RabbitMQ, ActiveMQ, ZeroMQ and IBM MQ.

## II. OVERVIEW

Modern messaging systems are built on message oriented middleware (MOM)[4] architecture that utilizes both the above mentioned paradigms- message queues and pub-sub systems. They provide an interface to connect distributed networks, hardware and applications. Distributed commit log technologies are one of the variants of MOM in which messages are committed in the form of logs. A log is a data structure where messages can get appended to. This is useful when message replay, guaranteed delivery and ordering play an important role in designing a messaging solution to integrate different applications together.

The popular commit log technologies include Apache Kafka, NATS Streaming, Google pub/sub, Microsoft event hubs and Amazon Kinesis. These frameworks that stream data are used for log metrics, website activity tracking, web services, enterprise applications, IOT and autonomous vehicles. Artificial intelligence which has taken the world by storm, involves a huge amount of data processing, which can only be addressed by a durable, well-structured data streaming framework. The following section provides a brief overview about NATS, Apache Kafka and Rabbit MQ followed by comparison of these three frameworks.

*A. NATS Streaming*

NATS is a lightweight, open-source cloud-based messaging system written in Go language that is open-source and is maintained by Synadia Group. It supports publisher-subscriber, request-reply and messaging queue models. The pub/sub system of NATS consists of publishers publishing messages to NATS subjects. Subscribers actively listening onto these subjects receive the messages. NATS server known as gnastsd provides for scalability by cutting off subscriptions if there is a timeout in connection to the server. Other features of NATS include clustered mode of servers and an always on dial tone for pub/sub system.

NATS streaming is a data streaming service for NATS server. The distinguishing feature between NATS and NATS streaming is that a NATS streaming server embeds a NATS server. NATS streaming API is used to communicate with the NATS server. Channels are the subjects in NATS Streaming in which clients receive data and producers send data to be put in message logs. NATS Streaming provides additional features such as at least once delivery of message, enhanced message protocol using Google protocol buffers, message persistence that's useful for message replay and durable subscriptions. Unfortunately, it does not support wildcard matching. Durable subscriptions essentially means that if a client were to restart, then the server will start delivery with the earliest message that's unacknowledged by that subscriber.

*B. Apache Kafka*

Kafka is a publish subscribe messaging system, written in Scala language, that is scalable, durable and fault tolerant. It is used by top companies such as LinkedIn, yahoo, twitter and others. The major use of Kafka is to stream data for real time analytics but is also used for monitoring, message replay, log aggregation, error recovery and website activity tracking. Kafka is simple to use and provides a high throughput and robust replication feature. In [5] Kafka used for Big data streaming demonstrates a pull-based model for clients to obtain data anytime and results in a high throughput. Kafka architecture comprises of producers, consumers, brokers, zookeeper, logs, partitions, records and topics [6]. Records have values and timestamp. Topics are categories for streams of records. Topics have logs that form storage on disks. Topic logs are divided into partitions that store records one after the other numbered by offset. These partitions are distributed over many brokers for high throughput. Kafka servers are known as brokers and many such form a Kafka cluster. Producers generate the streams of records that is put onto topics and consumers are responsible to subscribe to topics of their interest and retrieve the records using the offsets. Kafka can have consumer groups that help in load balancing. Many consumers in a consumer group can parallely retrieve data from different partitions of a topic. Kafka producer API and Kafka consumer API are used to carry out this process. A zookeeper is responsible to coordinate all the brokers in a cluster.

*C. RabbitMQ*

A message broker is an application that acts as an intermediary or a middleware for various applications. RabbitMQ is one such message broker system. It is open-sourced and incorporates Advanced Message Queuing Protocol (AMQP). It enables seamless asynchronous message-based communications between applications. The messages transported are loosely coupled, i.e. the sender and the receiver systems need not be up and running at the same time.

RabbitMQ is language agnostic [7]. It can be deployed and used across various operating systems. It supports languages like - .NET, Python, PHP, Ruby etc. It is written in Erlang Programming language. It is lightweight and can be deployed on cloud.

Messages are communicated over TCP connections. Major components involved in RabbitMQ are Publisher, Consumers, Exchange and Queues. A binding is a "link" or a rule that decides the route of a message to a particular queue. Every message that gets published in queue contains a payload and a routing key. The routing key decides the exact queue where a message needs to be delivered.

III. FEATURE COMPARISON

The three technologies described above are popular and provide competing similarities that it becomes difficult to choose the right framework. [8],[9] provide an in-depth analysis of Quality of Service (QOS) for different pub/sub systems that is useful to determine the scalability and quality of a system to cater to the growing consumer needs and adapt to the dynamic nature of communication. A detailed comparison is made between all three on different QoS factors.

*A. Message delivery*

Message guarantee or delivery forms the crux of quality of service. 'At most once' is the delivery where the message may or may not get delivered. This provides a high throughput. 'Exactly once' delivery occurs when message is received only one time. This requires expensive computations. 'At least once' delivery happens when message is sent at least once but can also get sent multiple times. This is useful for recovery from failure. Kafka provides at-most once, exactly once and at-least once delivery [10]. RabbitMQ provides at most once and at least once delivery. NATS streaming provides at most once and at least once delivery [11].

TABLE I
Feature Comparison Table

| Features | Apache Kafka | RabbitMQ | NATS Streaming |
|---|---|---|---|
| **Language developed in** | Scala | Erlang | Go |
| **Started In** | 2011 | 2007 | 2015 |
| **Messaging models supported** | pub/sub, Message queue | pub/sub, Message queue | pub/sub, Message queue, request-reply |
| **Brokered/Brokerless** | Brokered | Brokered | Brokered |
| **Throughput** | High | Medium to High | High |
| **Latency** | Low | Low to Medium | High |
| **Protocols supported** | Binary over TCP | AMQP, STOMP, MQTT | Google Protocol Buffer |
| **Message size** | 1 MB max | 2 GiB | 1 MB max |
| **Message Delivery** | At most once, Exactly once, At least once delivery | At most once, At least once delivery | At most once, At least once delivery |
| **Languages supported** | About 17 languages | About 30 languages | Officially about 12 languages |
| **Message Ordering** | Yes | Yes | Yes |
| **Message Storage** | Disk | In-memory/disk | In-memory/disk |
| **Distributed Units** | Topics | Queues | Channels |
| **Used By** | LinkedIn, Netflix, Facebook, Twitter, Chase Bank | Mozilla, AT & T, Reddit | Baidu, Ericcson, HTC, VMware, Siemen |

*B. Message Persistence*

It is the ability to retain the messages so that they are available even after a broker restart. RabbitMQ has persistence as an option and can be stored in-memory or on disk [12]. Even if a queue is set as durable in RabbitMQ it does not ensure message persistence. It has to be set by developer explicitly. In case of Kafka, the log can be stored on disk for persistence along with a message retention period set. NATS Streaming provides message persistence either in memory or using flat files. This is a configurable option for NATS.

*C. Message Ordering*

For RabbitMQ, ordering is present within a queue. For multiple queue subscribers the ordering can be maintained by the use of consistent hash exchange. For Kafka, ordering is present within a partition. For global ordering, expensive configurations need to be set up that decreases the performance. NATS Streaming provides subscribers messages in the order they were published by a single publisher but does not guarantee order delivery in case of multiple publishers.

*D. Throughput*

Throughput is defined as the number of messages per unit time that can be sent transferred between producers and consumers. The experiments carried out in [6] concluded that Kafka has a significantly higher throughput compared to RabbitMQ. The reason for this can be attributed to the sequential disk write feature and page-level caching of Kafka. Besides this, RabbitMQ waits for acknowledgements (ACKs) for each message and does not do batch processing, thus decreasing throughput. Albeit, RabbitMQ has the feature to turn off ACKs. The experiments carried out in [13] support this conclusion. In [14] the benchmark tests carried out for Kafka and NATS streaming show a comparable throughput but [2] describe tests in which throughput decreases considerably when the message size is increased by a large value for NATS Streaming. This cause might be related to the indirect connection of NATS Streaming to the NATS server. Apache Kafka is more matured compared to NATS Streaming and is loaded with many features and is designed for high throughput and scalable applications. NATS Streaming is rapidly evolving with the latest April 2019 release and is adding extensive features for high performance cloud native applications. Further research needs to be carried out to document the variations in performance of Kafka and NATS Streaming.

*E. Latency*

The delay in process is termed as latency. Kafka and RabbitMQ are capable of providing low-latency. NATS streaming provides a high magnitude of latency when compared to Kafka and RabbitMQ [2]. This can partially be attributed to the fact that NATS streaming API connects to NATS client API to get in contact with the NATS server which induces delay due to the indirect path. The experiments carried out in [10,15] conclude that for at-most once and at least once delivery in RabbitMQ the latency does not vary much for medium load and in the case of Kafka the latency almost doubles for load increase in 'at least once' delivery. If Kafka were to access storage from disk then the latency would rise further.

*F. Availability*

It is the capability of a system to maximize its uptime. The system needs to provide fault tolerance for high availability (HA). In [16], experiments are conducted for replication of queues in RabbitMQ to determine the availability and scalability impact. It is observed that for a single queue, performance is the highest and as the queue is mirrored the performance is affected, but this is a tradeoff for better fault tolerance. Kafka can replicate messages and store them in multiple brokers, this replication factor can be set by the developer. This is useful to make the data available in case of machine failures. Kafka also has a zookeeper [17] that coordinates between the brokers, producers and consumers and chooses another broker if one fails. Kafka provides high performance along with replication. NATS Streaming supports clustering and fault tolerance mode for HA. In clustering mode, data is replicated onto different cluster nodes using Raft Consensus algorithm whereas in FT mode data is stored on shared storage. However, to choose between the two modes, one should be aware that clustering has performance overhead because an acknowledgement is sent when data is replicated on all nodes.

*G. Scalability*

It defines the adaptability of a system to cater to a growing number of tasks such as producers or consumers or messages. RabbitMQ supports clustering [18] where in many nodes act as a single message broker. This is useful to balance the workload and scale the system to handle large number of messages. Kafka was built from ground up as a horizontal scaling system. With the co-ordination by the zookeeper, adding and deleting brokers makes Kafka easy to scale. NATS Streaming provides the clustering mode, but this is only for HA. Scalability is not well supported by NATS, and can be achieved by using the scheduling mechanisms provided by Go language.

## IV. DISCUSSION

The feature comparisons are listed concisely in Table I. The table also lists some of the similar features between the three, that can be investigated further by comparison with other messaging frameworks. The feature comparisons discussed, should be taken into consideration when designing an enterprise solution or developing a distributed IT system. Kafka is more mature when it comes to distributed log systems and should be considered for real-time data analytics that requires high throughput and low latency. NATS Streaming is relatively new and is useful for lightweight applications. It has a small support community that is growing rapidly with the new releases. The releases consist of more and more features evolving it into a high-performance messaging framework that can compete with Kafka. Kafka can be used in cases which are not real-time data critical, that is the streaming is unaffected even if few messages are missed/lost. Such applications are website user tracking, social media reactions and internet advertisements. RabbitMQ acts like a general message broker system and can be used for complex routing using exchanges and queues. These special routing is required in designing Internet of Things applications that connect many sensors. With the ACK transaction guarantee of RabbitMQ, all messages are guaranteed to be sent over to consumers, this feature is useful for financial transactions.

## V. CONCLUSION AND FUTURE SCOPE

This paper provided an introduction to the distributed messaging systems and outlined publisher/subscriber and messaging queue paradigm. The paper provided a brief overview on the working of three popular messaging frameworks- Apache Kafka, NATS Streaming and RabbitMQ. Following which, feature comparisons where discussed that revolved around message persistence, message ordering, throughput, latency, scalability and availability. These parameters are important to decide the suitability of a framework for an application. Hence, reader is expected to refer this paper for a thorough understanding to choose the right framework. Non-adherence to these parameters will result in unanticipated expenses in the development of an application.

Messaging frameworks will be the backbone of the future technology as more and more advance silicon chips are produced in the market that can handle high processing. Future work would include diving deeper into the technology with their newer releases and comparing them with upcoming frameworks.

ACKNOWLEDGMENT

This work was supported by Electronics and Communication Department, R.V. College of Engineering. Special thanks to Dr.Geetha K.S. for her kindness and encouragement on this work.